%% file: arXiv_Binkley_Moonen_Assessing_2022.tex
\documentclass[conference]{IEEEtran}
\usepackage[scaled=0.85]{helvet}
\usepackage{microtype}
\usepackage{amsmath} %
\usepackage{alg}
\usepackage{url}
\usepackage{listings}
\usepackage{graphicx}
\usepackage{verbatim}
\usepackage[svgnames]{xcolor}
\usepackage{colortbl}
\usepackage{xspace}
\usepackage{textcomp}
\usepackage{thmtools} 
\usepackage{thm-restate}
\usepackage{paralist}
\usepackage{booktabs} 
\usepackage{tikz}
\usepackage{balance}
\usepackage{xcolor}

\usepackage{cite}

\usepackage{pdfcomment}
\usepackage{hyperref}
\hypersetup{hidelinks}
\DeclareGraphicsExtensions{.pdf,.jpg,.png}

\lstdefinelanguage{nqC}{ %
  basicstyle=\small\sffamily,
  breaklines=true,
  showlines=false,
  showstringspaces=false,
  tabsize=2,
  escapechar=\&,
  columns=fullflexible,
  xleftmargin=.5in,
  numbers=left,
  classoffset=0,
  morekeywords={if,else,return},
  keywordstyle=\bfseries\color{bluekeywords},
  classoffset=1,
  morekeywords={int,double},
  keywordstyle=\bfseries\color{violet},
  classoffset=0,
  numberstyle=\tiny\color{black},
  numbersep=1em,
}

\lstdefinelanguage{WebAssembly}{
  basicstyle=\small\sffamily,
  breaklines=true,
  showlines=false,
  showstringspaces=false,
  tabsize=2,
  escapechar=\&,
  columns=fullflexible,
  xleftmargin=.5in,
  numbers=left,
  otherkeywords={},
  morekeywords=[1]{i32,f32,i64,f64},
  keywordstyle={[1]\bfseries\color{violet}},
  morekeywords=[2]{0},
  keywordstyle={[2]\color{violet}},
  morekeywords=[3]{add,const}
  keywordstyle={[3]\color{bluemunsell}},
  morekeywords=[4]{},
  keywordstyle={[4]\color{candypink}},
  morekeywords=[5]{module, func, param, mut, export, import, memory, data, get_local, set_local, elem, table, call,call_indirect, type, loop, block, br_if, if, br, end},
  keywordstyle={[5]\bfseries\color{bluekeywords}},
  morekeywords=[6]{=,;},
  keywordstyle={[6]\color{britishracinggreen}},
  morekeywords=[7]{(,),[,],.},
  keywordstyle={[7]\color{black}},
  numberstyle=\tiny\color{black},
  numbersep=1em,
  rulecolor=\color{black},
  morecomment=**[l][\itshape\color{greencomments}]{;;},
}

\newcommand*{\Comment}[1]{\hfill\makebox[5.0cm][l]{#1}}%

\newcommand{\WASMinl}[1]{``\WASMfrag{#1}''}
\newcommand{\Cinl}[1]{``\Cfrag{#1}''}
\newcommand{\WASMfrag}[1]{\lstinline[language=WebAssembly,basicstyle=\sffamily]{#1}}
\newcommand{\Cfrag}[1]{\lstinline[language=nqC,basicstyle=\sffamily]{#1}}

\newlength{\floatcorrection}
\newcommand{\head}[1]{\par\noindent\textbf{#1}}

\newcommand{\approach}{\textsc{\textit{n}VORBS}\xspace}
\newcommand{\eG}{$\mathcal{G}$\xspace}
\newcommand{\eC}{$\mathcal{C}$\xspace}
\newcommand{\eW}{$\mathcal{W}$\xspace}
\newcommand{\eGC}{$\mathcal{GC}$\xspace}
\newcommand{\eCW}{$\mathcal{CW}$\xspace}
\newcommand{\eGW}{$\mathcal{GW}$\xspace}
\newcommand{\eGCW}{$\mathcal{GCW}$\xspace}
\newcommand{\gcc}{\textsf{gcc}\xspace}
\newcommand{\clang}{\textsf{clang}\xspace}
\newcommand{\wasm}{\textsf{wasm}\xspace}

\title {Assessing the Impact of Execution Environment \\ on Observation-Based Slicing}

\author{
\IEEEauthorblockN{%
David Binkley\IEEEauthorrefmark{1} \quad\quad
Leon Moonen\IEEEauthorrefmark{4}\\[.8ex] 
}
 
\IEEEauthorblockA{\IEEEauthorrefmark{1}
  Loyola University Maryland, Baltimore, MD, USA --
  E-mail: binkley@cs.loyola.edu%
}

\IEEEauthorblockA{\IEEEauthorrefmark{4}
  Simula Research Laboratory, Oslo, Norway --
  E-mail: leon.moonen@computer.org%
}
}

\colorlet{bluekeywords}{blue}

\makeatletter
\def\ps@IEEEtitlepagestyle{%
  \def\@oddfoot{\mycopyrightnotice}%
  \def\@evenfoot{}%
}
\def\mycopyrightnotice{%
  \hspace*{3mm}\includegraphics[width=2cm]{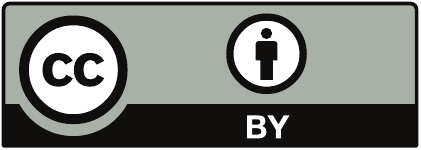}%
  \hspace*{2mm}\raisebox{2.5mm}{%
   	  \parbox{\columnwidth}{\footnotesize This work is licensed under a Creative Commons \\ Attribution 4.0 International (CC BY 4.0) license.}%
   	  \hspace*{-68pt}\mbox{1}\hspace{20pt}\fbox{\parbox{.9\columnwidth}{\footnotesize\textsl{Accepted for publication in the 22nd IEEE International Working Conference on Source Code Analysis and Manipulation (SCAM 2022).}}}%
  }%
  \gdef\mycopyrightnotice{}%
}
\makeatother

\begin{document}

\pagestyle{plain}

\maketitle

\begin{abstract}

Program slicing reduces a program to a smaller version that retains a chosen computation, 
referred to as a \emph{slicing criterion}.
One recent multi-lingual slicing approach, \emph{observation-based slicing (ORBS)},
speculatively deletes parts of the program and then %
executes the code.  
If the behavior of the slicing criteria is unchanged, the speculative deletion is made permanent.

While this makes ORBS language agnostic, it can lead to the production of some
non-intuitive slices.
One particular challenge is when the \emph{execution environment} plays a role.
For example, ORBS will delete the line \Cinl{a = 0} if the memory location assigned to \Cfrag{a} contains zero before executing the statement, 
since deletion will not affect the value of \Cfrag{a} and thus the slicing criterion.
Consequently, slices can differ between execution environments due to factors such as initialization and call stack reuse.

The technique considered, \approach, 
attempts to ameliorate this problem by \emph{validating} a candidate slice in \textit{n} different execution environments.
We conduct an empirical study to collect initial insights into how often the execution environment leads to slice differences.  
Specifically, we compare and contrast the slices produced by seven different instantiations of \approach.
Looking forward, the technique can be seen as a variation on metamorphic testing, 
and thus suggests how ideas from metamorphic testing might be used to improve dynamic program analysis.
\end{abstract}

\section{Introduction} 

\noindent
Program slicing is a program decomposition technique that has a wide range of
applications in various areas such as
 debugging,
 program comprehension,
 software maintenance,
 refactoring,
 testing,
 reverse engineering,
 tierless or multi-tier programming,
 commit decomposition, and
 vulnerability detection~\cite{weiser:slicing81, gallagher1991, dbmh:advances, silva-survey:12, 2018:programming:philips, 
2018:scam:muylaert, vulslicer:20}.
At its introduction, Weiser defined program slicing as follows:
``Starting from a subset of a program's behavior, slicing \emph{reduces} that
program to a minimal form which still produces that behavior''~\cite{weiser:slicing81}.
This behavioral subset is referred to as a \emph{slicing criterion}.
Surprisingly, most slicing algorithms try to decide which code
should be \emph{retained} to preserve the criterion.

In contrast, \emph{observation-based slicing (ORBS)}~\cite{binkley2014orbs} 
is closer to
Weiser's definition.
Rather than using static or dynamic analysis to find which parts of
the program to include in a slice, ORBS tentatively removes parts from
the program and \emph{observes} the impact of their removal.  
Removals that do not impact the behavior of interest are made permanent.
This enables ORBS to easily slice multi-lingual programs~\cite{binkley2014orbs}, 
programs with non-standard semantics~\cite{jss17-vorbs:Yoo}, and handle
``hidden'' dependences such as those caused by reading and writing a common file~\cite{Binkley:2015scam}.

For all its virtues, ORBS has two significant drawbacks.  
First, it requires considerable computational effort to compute a slice.
Second, real-world execution environments can lead to unexpected behavior. 
The first of these can be mitigated by applying greater computational power
and through (non-trivial) engineering work~\cite{islam2016porbs}.
In some ways, the second drawback poses the greater challenge and is the topic
of this paper.

\begin{figure}
\vspace{1ex}
\begin{lstlisting}[language=nqC,xleftmargin=1.7em,lineskip=-0.1pt,frame=lines]
  int f() {
    int a;
    a = 42;
    return a;
  } 
  int g() {
    int b;
    b = 42; & \label{ex1-deletable} \Comment{// Is this statement deletable?} &
    return b;
  }
  main() {
    int x, y;
    x = f(); & \label{ex1-before-criterion} &
    y = g(); & \label{ex1-criterion} \Comment{// Slice here on final value of y.} &
  }
\end{lstlisting}
\vspace{-2.0ex}
\caption{An example hidden dependence cause by memory location reuse.}
\label{fig:stack-overlap}
\vspace{-3ex}
\end{figure}

In theory (as opposed to in practice), this challenge does not exist.
It is possible to construct a well-defined formal semantics for a programing
language such that ORBS can maintain execution behavior while cleanly removing
unnecessary code.  
Doing so in practice is less straightforward due to choices made in real-world
compilers and runtime environments.
For example,
consider the code in Figure~\ref{fig:stack-overlap}.
All (correct) dependency-based slicers will (correctly) conclude that the
assignment on Line~\ref{ex1-deletable} is in the slice taken with respect to
\textsf{y} at Line~\ref{ex1-criterion} because of the transitive data dependence of \textsf{y}'s
value on the assignment to \textsf{b}.
The same would be true of ORBS using an idealized execution environment. 
However, on many systems, stack activation records are reused, and thus 
\textsf{a} and \textsf{b} \emph{can} share the same location in memory, which
leaves \textsf{b} holding the correct value even if Line~\ref{ex1-deletable} is omitted from the slice (assuming no well-timed interrupt
changes the stack).

\head{Contributions:} 
To mitigate the impact of a particular execution environment, this paper proposes \approach, 
an ORBS variant that validates candidate slices in \textit{n} different execution environments.
The idea is that the weaknesses of one execution environment are covered by at
least one of the others, 
so if \textit{n} is sufficiently large and a candidate slice behaves similarly in all \textit{n} environments, we have evidence that it is a correct slice.

To gain some initial insight into how often the execution environment leads to
the problems described above, we conduct an empirical study in which we compare and
contrast the slices produced by seven different instantiations of \approach. 

\section{Observation-Based Slicing}
\label{sec:slicing}

\noindent
ORBS computes \emph{dynamic backward executable slices}~\cite{silva-survey:12}. 
To do so, it repeatedly attempts to delete a window of one to four consecutive lines.  
Using a window of one to four lines has been empirically shown to balance the number of lines
that can be deleted in a single step versus the time wasted on larger deletions
that more often than not change the program's behavior~\cite{binkley2014orbs}.
In greater detail, 
for the slice taken with respect to variable $v$ at program location (line) $l$,
ORBS first instruments the program to print the value of $v$ immediately after $l$.
Next, the program is executed on its test suite, and the printed values of $v$
are recorded as an oracle.
ORBS then repeats its main loop where it iterates over the program,
speculatively deleting a window of consecutive lines.
The resulting program is compiled and executed or interpreted in an
\emph{execution environment}, and if its output matches the oracle, 
then the speculative deletion is made permanent.
Finding motivation in metamorphic testing, \approach extends ORBS by checking
if the output matches the oracle in \textit{n} execution environments.

\section{Experiment Design}

\begin{figure}[b]
\vspace{-2.0ex}
\centering
\newcommand{\dist}{.6cm}
\resizebox{0.35\columnwidth}{1in}{%
\begin{tikzpicture}[node distance=1.5cm,line width=1pt]
\title{FYI ME $A_4$}

\node(GCW)                     {\eGCW};
\node(GW)      [below of=GCW]  {\eGW};
\node(GC)      [left of=GW]   {\eGC};
\node(CW)      [right of=GW]    {\eCW};

\node(C)      [below of=GW]      {\eC};
\node(W)      [right of=C]    {\eW};
\node(G)      [left of=C]     {\eG};

\draw(GCW) -- (GC);
\draw(GCW) -- (CW);
\draw(GCW) -- (GW);

\draw(GC) -- (G);
\draw(CW) -- (W);
\draw(GW) -- (G);
\draw(GC) -- (C);
\draw(CW) -- (C);
\draw(GW) -- (W);

\end{tikzpicture}%
}%
\vspace{-1.0ex}%
\caption{Containment lattice of the seven instantiations.}
\label{fig:lattice}
\end{figure}
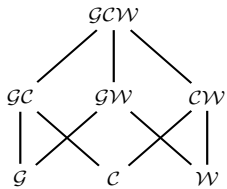

\noindent
To investigate the impact of the execution environment on observation-based slicing,
we study seven instantiations of \approach using three execution environments, as shown in Figure~\ref{fig:lattice}.
All seven are identical except for the compilers used and the runtimes in which
programs are executed.
The bottom three instantiations \eG, \eC, and \eW use a single execution environment 
(respectively based on a selection of modern \textsf{C} compilers \gcc, \clang, and \clang using the option \textsf{--target=wasm32-unknown-wasi}, 
which we refer to as \wasm in the following).
Both \eG and \eC build an executable that is run directly from the operating system.
\eW builds a WebAssembly binary that is executed in the \textsf{wasmer} virtual environment.

The middle three use pairs of execution environments.
Specifically, \eGC uses both \gcc and \clang, \eCW uses \clang and \wasm, and 
\eGW uses \gcc and \wasm.
In each instantiation, both executables are built, and the output
of each is checked against the oracle.
Finally, \eGCW 
uses
all three execution environments.
\eGCW is expected to be the most strict (i.e., accept fewer deletions) and thus produce larger slices.
It is also expected to be the least susceptible to unwanted slicing behavior.

\section{Research Questions}

\head{RQ1: How often does the execution environment impact the slice?}
In an ideal world, all execution environments would produce the same slice.
This question empirically investigates how close reality comes to this ideal.

\head{RQ2: What containment relations exist between the slices produced by the different instantiations?}
Intuitively \approach instantiations that involve stricter requirements (i.e., with larger \textit{n}) should produce larger
slices because they can delete fewer lines.
Thus we expect using both $gcc$ and $clang$ to produce larger slices than using only one of the two.

\head{RQ3: What qualitative patterns are found in the slices?}
How does the source code of slices produced by the seven instantiations for the same criterion and program compare?
For example, \gcc and \clang are similar, so one might expect them to produce
similar slices.
As a second example, \wasm's use of a virtual machine might be more stable
because each execution starts with the virtual machine in an identical initial
state.

\section{Evaluation}

\noindent This section describes the subjects used in the empirical study,
as well as the execution environments and hardware used to run the experiments.
Next, we address the three research questions.

\head{Subjects:}
The experiments consider the 69 \textsf{C} programs shown in
Table~\ref{tab:subjects}, which were used in prior program analysis research: 

\begin{itemize}
  \item Programs from the slicing literature: a variation on the original example of
Weiser (\textsf{sumprod})~\cite{weiser:slicing81}, the SCAM Mug example (\textsf{scam})~\cite{ward2003}, the Montr{\'e}al Boat Example (\textsf{mbe})~\cite{danicic2002}, and word count (\textsf{wc})~\cite{gallagher1991}.
  \item Programs from the M{\"a}lardalen WCET benchmark for comparing and evaluating WCET analysis tools~\cite{wcet}.
  \item Programs from the \emph{Benchmarks Game}~\cite{benchmarksgame}, which are designed to benchmark language implementations.
  \item Two multi-file programs (\textsf{bc} and \textsf{indent}) used in previous slicing studies~\cite{dbngmh:empirical}.  
\end{itemize}

\begin{table}
 \caption{Subject Systems.  For Multi-file programs the sliced file is shown in parenthesis. 
(``\textsf{rc}''' abbreviates \textsf{reverse-compliment)}}
\centering
\label{tab:subjects}
\vspace*{-.5ex}
\begin{tabular}{l @{~~~} r @{~~~}r}
Program  & SLoC & Slices \\
\midrule
\textsf{adpcm}& 585 & 168 \\
\textsf{bc (bc.c)} & 8\,594 & 54 \\
\textsf{bc (execute.c)} & 9\,140 & 88 \\
\textsf{binary-trees1}& 91 & 14 \\
\textsf{bs}& 46 & 9 \\
\textsf{bsort100}& 61 & 8 \\
\textsf{cnt}& 78 & 17 \\
\textsf{compress}& 357 & 74 \\
\textsf{cover}& 625 & 199 \\
\textsf{crc}& 94 & 16 \\
\textsf{duff}& 44 & 1 \\
\textsf{edn}& 170 & 42 \\
\textsf{expint}& 73 & 23 \\
\textsf{fac}& 25 & 3 \\
\textsf{fankuchredux1}& 79 & 11 \\
\textsf{fankuchredux5}& 115 & 20 \\
\textsf{fasta1}& 126 & 19 \\
\textsf{fasta2}& 264 & 34 \\
\textsf{fasta3}& 90 & 6 \\
\textsf{fasta5}& 111 & 15 \\
\textsf{fasta7}& 231 & 32 \\
\textsf{fasta8}& 150 & 14 \\
\textsf{fasta9}& 163 & 16 \\
\textsf{fdct}& 138 & 86 \\
\textsf{fft1}& 128 & 41 \\
\textsf{fibcall}& 27 & 8 \\
\textsf{fir}& 54 & 15 \\
\textsf{hanoi\_c}& 178 & 21 \\
\textsf{indent (indent.c)} & 6\,680 & 223 \\
\textsf{indent (parse.c)} & 5\,213 & 28 \\
\textsf{insertsort}& 33 & 5 \\
\textsf{janne\_complex}& 38 & 7 \\
\textsf{jfdctint}& 119 & 65 \\
\textsf{lcdnum}& 62 & 5 \\
\textsf{lms}& 172 & 51 \\
\\
\end{tabular} ~~
\begin{tabular}{l @{~~~} r @{~~~}r}
Program  & SLoC & Slices \\
\midrule
\textsf{ludcmp}& 109 & 25 \\
\textsf{mandelbrot2}& 66 & 18 \\
\textsf{mandelbrot9}& 66 & 7 \\
\textsf{matmult}& 55 & 7 \\
\textsf{mbe}& 33 & 12 \\
\textsf{minver}& 201 & 49 \\
\textsf{nbody1}& 92 & 13 \\
\textsf{nbody2}& 107 & 14 \\
\textsf{nbody3}& 90 & 20 \\
\textsf{nbody6}& 93 & 13 \\
\textsf{nbody7}& 137 & 23 \\
\textsf{ndes}& 196 & 39 \\
\textsf{ns}& 31 & 4 \\
\textsf{prime}& 51 & 1 \\
\textsf{printtokens}& 570 & 81 \\
\textsf{printtokens2}& 408 & 75 \\
\textsf{qsort-exam}& 124 & 22 \\
\textsf{qurt}& 120 & 16 \\
\textsf{rc-5}& 83 & 11 \\
\textsf{rc-6}& 96 & 15 \\
\textsf{replace}& 542 & 309 \\
\textsf{scam}& 35 & 16 \\
\textsf{schedule2}& 292 & 74 \\
\textsf{schedule}& 314 & 58 \\
\textsf{select}& 131 & 22 \\
\textsf{spectral-norm1}& 57 & 10 \\
\textsf{st}& 98 & 14 \\
\textsf{statemate}& 1\,354 & 364 \\
\textsf{sumprod}& 17 & 8 \\
\textsf{tcas}& 142 & 43 \\
\textsf{totinfo}& 348 & 54 \\
\textsf{triangle}& 59 & 7 \\
\textsf{ud}& 81 & 22 \\
\textsf{wc}& 49 & 17 \\
\\
Total 
& 40\,311 & 2\,921 \\
\end{tabular}%
\vspace*{-3.7ex}%
\end{table}

\noindent
We omit programs from the second and third sources that span multiple files,
that fail to compile with \textsf{-lm} as the only compiler flag enabled, and that
are unsupported by the \textsf{pycparser} Python library.
We use \textsf{pycparser} to normalize (i.e., pretty-print) the code and to
instrument it (i.e., add a \Cfrag{printf} statement that captures the criterion).
For multi-file programs, ORBS slices a specified file from the program (this
file is show in Table~\ref{tab:subjects} within parenthesis).
Thus there is no practical limit on the size of the program ORBS can slice.

\head{Execution Environments:}
The three execution environments respectively use \textsf{gcc v12.0.0
20210720} running the compiled program natively, 
\textsf{clang v13.0.0 (50302feb)} also running the compiled program natively,
and \clang compiling to WebAssembly and using the \textsf{wasmer 2.0.0} WebAssembly runtime.

\head{Hardware:}
All data was generated using a 676-core computing cluster using 2.20GHz Xeon(R) E5-2650 CPUs.  
The cluster has 256GB RAM per node and terabytes of HDD and fast SSD storage
all connected using a 56 gigabit Infiniband network.

\head{RQ1:}
To answer how often the execution environment impacts the slice, we
compare the individual slices produced by each pair of \approach instantiations for the same program and slicing criterion.
With seven instantiations and 2921 slices, this leads to ((7 * 6) / 2) * 2921 = 61\,341 comparisons.
In 47\,269 (77\%) of these, both instantiations produce the same slice.
This is encouraging.
It means that three-quarters of the time, the execution environment does not impact the slice.
Of the remaining 14\,072 comparisons, 
in 9\,279 (66\%) cases one slice includes a subset of lines from the other. 
The remaining 4\,793 cases are different slices (neither is a subset of
the other).

Some slices may exhibit non-deterministic behavior in certain environments.
For example, after removing Line~\ref{ex1-deletable} of Figure~\ref{fig:stack-overlap},
if an interrupt occurs between the execution of Lines~\ref{ex1-before-criterion} and \ref{ex1-criterion}, 
and this interrupt changes the stack, 
then \textsf{y} may no longer be assigned the value 42.
Because it is possible that these slices degrade the analysis, 
we remove from all instantiations those slices that exhibit non-deterministic behavior in any one of them.
This filter turns out to have minimal impact, removing only 22 slices.
Post removal there are 60\,879 comparisons of which 46,954 (77\%) are identical, 
and of the remaining 13\,925 comparisons, 9\,142 (66\%) are subset relations.

The second filter removes slices where the criterion is not part of the
slice in all seven instantiations.
This typically occurs because of an inadequate test suite.
This filter has a greater impact, removing 1\,118 slices.
Post removal there are 37\,863 comparisons of which 24\,907 (66\%) are
identical, and of the remaining 12\,956 comparisons, 8\,269 (64\%) are subset
relations. \linebreak
\hspace*{\parindent} 
Finally, applying both filters together leaves 1781 slices, yielding 37\,401 comparisons
of which 24\,592 (66\%) are identical, and of the remaining 12\,809 comparisons
8\,134 (64\%) are subset relations.

In summary for RQ1, the execution environment affects the slice in roughly one of three cases.
Note that this is a conservative result because some differences are very minor.

\begin{table}
\caption{Subset relations for related environments.}
\label{tab:subsets}
\centering\setlength{\tabcolsep}{10pt}
\begin{tabular}{l r r r r}

comparison & $=$ & $\supset$ & $\subset$ & $\neq$ \\ \midrule

\eG vs.\ \eGC   &  1278 &   3   &  328 &  172 \\
\eG vs.\ \eGW   &  1013 &   4   &  658 &  106 \\
\eG vs.\ \eGCW  &   981 &   0   &  622 &  178 \\

\midrule
\eC vs.\ \eGC   &  1432 &   9   &  214 &  126 \\
\eC vs.\ \eGW   &  1154 &   3   &  471 &  153 \\
\eC vs.\ \eGCW  &  1096 &   4   &  488 &  193 \\

\midrule
\eW vs.\ \eGW   &  1243 &  14   &  289 &  235 \\
\eW vs.\ \eCW   &  1276 &   4   &  273 &  228 \\
\eW vs.\ \eGCW  &  1206 &   3   &  309 &  263 \\

\midrule
\eGC vs.\ \eGCW &  1178 &   1   &  446 &  156 \\
\eGW vs.\ \eGCW &  1558 &   3   &   92 &  128 \\
\eCW vs.\ \eGCW &  1483 &   9   &  191 &   98 \\

\end{tabular}%
\vspace*{-3ex}%
\end{table}

\head{RQ2:} The seven \approach instantiations validate speculative deletions
using one or more execution environments.
RQ1 shows that the execution environment affects roughly one in three slices.
For these slices, validating in multiple execution environments can be
expected to restrict the number of successful deletions.
RQ2 investigates the impact this restriction has by analyzing the containment
relations between the slices produced by the seven \approach instantiations.

Starting with the 1781 slices of the doubly filtered data created while addressing RQ1, 
Table~\ref{tab:subsets} presents the comparisons in groups of three.
The first three groups compare the slices of the singleton instantiations \eG,
\eC, and \eW with those instantiations that involve additional systems.
For example, the first group compares \eG with the three others that involve
\gcc: \eGC, \eGW, \eGCW.
The first column shows the comparison, while the second counts the number of slices 
where \eG alone produces the same slice as \eGC, \eGW, and \eGCW: 
1278 (72\%), 1013 (57\%), and 981 (55\%), respectively.
The third column shows how often slices produced by \eG are a superset of (i.e.,
larger than) the corresponding slice produced by the other instantiations
(this is unexpected because \eG alone is less restrictive).
The fourth column shows how often the slice produced by \eG is a subset of the
other (the expected outcome), 
and the fifth column shows the number of slices where neither is a subset of the other.
The first group shows the expected pattern where \eG is a superset in only a few
cases (specifically 3, 4, and 0), while it is a subset in respectively 328,
658, and 622 cases.

The following two groups repeat these comparisons for respectively \eC and \eW.
Observe that the overall pattern is the same.
It is interesting to note that \eG, which uses \gcc, produces the fewest
supersets and by far the most subset relations.

The last group in Table~\ref{tab:subsets} compares 
the three instantiations that use two environments (i.e., \eCW, \eGW, and \eCW) 
with the instantiation that uses all three (i.e., \eGCW).
These comparisons show evidence that using more environments increases stability.
For example, an average of 79\% of the slices in the last group are identical
compared to only 66\% averaged over the nine singleton comparisons.
Finally, the individual rows show evidence that \wasm accounts for much of the greater similarity. 

In summary for RQ2, while not universal, the subset relations support the
intuition that the stricter requirements of validating slices in multiple
execution environments produce larger slices because fewer lines can be deleted.

\head{RQ3:} RQ3 takes a qualitative look at the slices of the filtered data 
to focus on differences in the slices themselves.
While it is not practical to compare all slices by hand, 
it is possible to do this for some of the smaller programs.
This section presents several interesting examples that are short enough
to explain in limited space.
To begin with, for 18 of the 69 test subjects, all seven instantiations produce
the same set of slices.

Looking at the individual slices, we first consider the slice of \textsf{wc}
taken with respect to the computation of \textsf{inword}, which is
\textsf{true} when the current character is part of a word.
The variable \textsf{inword} is initialized 
by the statement \Cinl{inword = 0}.
When using \clang, this initialization %
is retained.
In contrast, when using \gcc and \wasm, it is deleted because the memory
location assigned to \textsf{inword} happens to initially hold the value zero.

As a related example, \eC (and \eCW, \eGC, and \eGCW)   %
retain the declaration \Cinl{int lines} in the slice taken with respect to
the final value of \textsf{words}, even though \textsf{lines} is not used
elsewhere in the code.
To understand why, the first step is to note that a previous pass deleted
\Cinl{words = 0} for the same reason that \textsf{inword} was deleted above.
However, removing \Cinl{int lines} in a subsequent pass
changes the address of \textsf{words} to one with a non-zero
value, thereby preventing the removal of the declaration.
That two of the three execution environments retain this statement suggests 
that future work might consider voting instead of requiring equal
slices in all environments.

Compiling WebAssembly puts certain constraints on code.
For example, it requires that the effect on the stack must be well-typed.
Each WebAssembly instruction has a specific \emph{stack type} $t_1^* \rightarrow t_2^*$,
where $t_1^*$ is the expected sequence of types for the values on top of the
stack before execution and $t_2^*$ is the sequence of
types for the values on top of the stack after execution.
For instance, the wasm instruction \WASMinl{i32.const 42} has type $~~\rightarrow$
\WASMfrag{i32}'', meaning that it does not need anything from the stack and pushes
one value of type \WASMfrag{i32}.
These constraints result in \emph{surprising} compilation choices when \wasm
compiles an \Cfrag{if} where ORBS is attempting to delete the \Cfrag{else} branch of the code.
Consider the following function with conditional code:
\begin{lstlisting}[language=nqC]
  int ishappy(...) { & \Comment{// returns a boolean using an int}&
    if (Cond)
      return 0;
    else          & \Comment{// ORBS attempts to delete }&
      return 1;   & \Comment{// these two lines}&
  }
\end{lstlisting}
When \gcc compiles this code without the else branch, it moves a zero into the return
location when \Cfrag{Cond} is true, otherwise it does nothing, and thus the
value in the return location is unchanged.   
Commonly this value is non-zero, which \textsf{C} programs interpret as \Cfrag{true}.
In the \wasm case, because of its well-typed stack requirement in the compiled code, 
both branches of an \WASMfrag{if} statement must have the same effect on the stack.
What the \wasm compiler does when presented with the above code without Lines 4 and 5 is surprising, 
yet legal and satisfies the requirement:
it replaces the \WASMfrag{if} statement altogether by the compiled version of \Cinl{return 0}.

A second \wasm example relates to the code in Figure~\ref{fig:stack-overlap}.
When excluding the code on Line~\ref{ex1-deletable}, the compiler generates:
\begin{lstlisting}[language=WebAssembly]
 (func $g (type 2) (result i32)
    (local $0 i32)
    (get_local $0)
 ) 
\end{lstlisting}
When executing this in the \textsf{wasmer} runtime, it evaluates to zero (most likely because 
the runtime initializes the memory it uses on startup). 
However, the oracle expects 42, 
which means that ORBS using \wasm retains Line~\ref{ex1-deletable} in the slice.

Next consider the slice of \textsf{totinfo} with respect to \textsf{n} at
Line 199, which includes the following code \emph{except} in the \eW instantiation:
\begin{lstlisting}[language=nqC,firstnumber=79]
  if ( r * c > MAXTBL ) {   
    return EXIT_FAILURE;
  }
\end{lstlisting}
\noindent
If this test is omitted, the executables produced by \gcc and \clang both report
a memory violation.
However \wasm allocates sufficient memory that the subsequent out of bounds
array accesses do \emph{not} generate a memory violation, and thus using \eW the statement
can be deleted.
(Interestingly, running the \gcc produced executable in \textsf{gdb} does not
generate a fault.)

Finally, a compiler difference can be observed in the slice of \textsf{totinfo} 
with respect to \textsf{i} at Line 389:
for this slice, \clang prevents ORBS from removing the declaration and return of
the variable \textsf{info}, while \gcc permits their removal.
\begin{lstlisting}[language=nqC, firstnumber=303]
    double info;   & \Comment{/* accumulates information measure */}
\end{lstlisting}
\vspace{-1em}
\hspace{0.60in}\Cfrag{...}
\vspace{-0.5em}
\begin{lstlisting}[language=nqC, firstnumber=414]
  ret3:
    return info;
  }
\end{lstlisting}
\noindent
The root cause of this difference is a rare compiler disagreement on the definition of correct C syntax, 
which deserves further investigation.
Using \gcc, the return (and subsequently the declaration of \textsf{info}) can be excluded without problem
(even when using the stricter \textsf{--ansi} flag).
In contrast, removing the return causes \clang to issue a syntax error that a statement is expected but missing.
Because it cannot remove the return, \clang must also retain the declaration.
While ORBS cannot do this, if the return is replaced by \Cinl{return 42}
\clang produces the expected oracle output and can remove the declaration.

In summary for RQ3, the most common pattern seen occurs when a variable
initialization can be removed without changing the behavior of the program.
In addition, the WebAssembly stack validation requirement has the interesting
effect of avoiding the non-deterministic return value found when removing a
\Cfrag{return} statement found in an \Cfrag{else} branch.

\head{Threats to validity:}
In addition to the standard internal threats found when using tools an analyze 
software, threats to external validity exist when asking if our results extend
to other environments, other programs, and especially other programming languages. 
For example, we studied relatively small \textsf{C} programs from the program analysis literature.
Although these are typically aimed at covering all language features of interest, 
we do not know to what extent our results generalize to large systems or, for example, embedded \textsf{C} code.
Moreover, the language \textsf{C} has rather ``loose'' semantics,
so other languages that are more completely and precisely defined would likely
yield more consistent results.
One specific threat follows from using MD5 hashes to compare the result of executing a slice with the oracle.
There is the potential for errant matches when doing more than $2^{64}$ comparisons.
This can be easily addressed by using a longer hash, such as SHA256.

\section{Related Work}
\label{sec:related}

\noindent
The most closely related work to ours is that related to ORBS~\cite{binkley2014orbs}, 
which includes the parallel implementation~\cite{islam2016porbs} used to implement \approach.
The modification of the compile and runtime system was to some degree inspired
by the work on slicing languages with non-standard semantics such as picture description languages~\cite{jss17-vorbs:Yoo}.
Finally, the use of \wasm finds its roots in the slicing of WebAssembly~\cite{slicing_wasm:icse-2022}.  

Traditional software testing relies on the presence of a \emph{test oracle}
that decides what is the correct output or expected behavior for a given test input.  
Automated software testing uses the oracle to identify failures, i.e., when the behavior of the software system deviates from the oracle.  
When a system is `non-testable'~\cite{weyuker81}, because an oracle is unavailable, 
or prohibitively expensive to use, \emph{metamorphic testing} provides a way forward~\cite{segura2016:survey}.
It is based on defining one or more relations between the outputs of a program that must hold for a series of inputs.
A typical example is the sinus function for which the following metamorphic relation holds for any $x$: $sin(x) = sin(\pi - x)$.
Another example is a cryptographic system that supports both stream and block modes.
Metamorphic testing would encrypt a message using the two modes and expect the same encrypted message.
Likewise, when testing autonomous vehicle software where producing an oracle is expensive, 
metamorphic testing can test car behavior on a set of transformed images that should result in the same behavior~\cite{testing-auto:ieee21}.

Closer to our use case of ensuring that slices behave the same in different execution environments, 
metamorphic testing has been used to ensure correct behavior of Datalog engines~\cite{mansur2021:metamorphic},
find bugs in the implementation of compilers~\cite{le2014:compiler}, 
as well as increase the validity of various simulation models~\cite{olsen2019:increasing}.

\section{Concluding Remarks}
\label{sec:conc}

\head{Contributions:}
To mitigate any negative impact of a particular execution environment on observation-based slicing, 
we propose \approach, an ORBS variant that validates candidate slices using \textit{n} different execution environments.
By considering seven differing instantiations of \approach, 
we empirically investigate how widespread the impact of the execution environment is observation-based slicing in practice.

Our data suggest that the environment affected slices in roughly one of three cases. 
Thus, validating the correctness of slices in multiple environments helps to cover weaknesses in one of the environments, 
and increases the evidence that only correct slices are produced.
The data also shows that validating slices in multiple environments results in larger slices. 
Finally, our qualitative analysis uncovers several interesting findings, 
such as the value that WebAssembly's more structured runtime brings to dynamic analyses such as ORBS.

\head{Future Work:}
Further empirical work is needed to evaluate the benefits and drawbacks of \approach.
The qualitative examples suggest some interesting directions for future work, such
as the use of a voting scheme when using multiple environments.
In addition, there are several opportunities for efficiency
improvements. 
For example, a parallel algorithm can attempt the deletion of all window
sizes concurrently and retain the largest successful deletion~\cite{islam2016porbs}.
In addition, the validation in multiple execution environments can be
easily parallelized. 
Finally, one can see \approach as a variation on metamorphic testing.
Looking forward, the use of metamorphic testing ideas might find broader use in
dynamic source code analysis. 

\medskip

\head{Acknowledgements:}
This work was supported by NSF grant \#1626262, which provided the
cluster used to generate the empirical data.
Leon Moonen was supported by the Research Council of Norway through the secureIT project (\#288787).

\IEEEtriggercmd{\newpage\vspace*{2ex}}
\bibliographystyle{IEEEtran} 
\input{arXiv_Binkley_Moonen_Assessing_2022.bbl}

\end{document}